\documentclass[a4paper,12pt]{article}
\usepackage{graphicx}
\usepackage[american]{babel}

\begin{document}
\title{Mandelbrot set in coupled logistic maps and in an electronic experiment}
\author{Olga~B.~Isaeva$^{1,2}$, Sergey~P.~Kuznetsov$^{1,2}$,
Vladimir~I.~Ponomarenko$^1$}
\date{}
\maketitle\begin{center} \emph{$^1$ Institute of Radio-Engineering
and Electronics of RAS, \\ Zelenaya 38, Saratov, 410019,
Russia}\end{center}

\maketitle\begin{center} \emph{$^2$ Saratov State University, \\
Astrakhanskaya 83, Saratov, 410026, Russia}\end{center}

\maketitle
\begin{abstract}

We suggest an approach to constructing physical systems with
dynamical characteristics of the complex analytic iterative maps.
The idea follows from a simple notion that the complex quadratic
map by a variable change may be transformed into a set of two
identical real one-dimensional quadratic maps with a particular
coupling. Hence, dynamical behavior of similar nature may occur in
coupled dissipative nonlinear systems, which relate to the
Feigenbaum universality class. To substantiate the feasibility of
this concept, we consider an electronic system, which exhibits
dynamical phenomena intrinsic to complex analytic maps.
Experimental results are presented, providing the Mandelbrot set
in the parameter plane of this physical system.
\end{abstract}

\maketitle PACS number(s): 05.45.-a, 05.45.Df, 05.45.Xt

\newpage

One of the rich and fascinating sub-disciplines in nonlinear
dynamics is the theory of iterative complex analytic mappings. A
well-known example is the quadratic map
\begin{equation} \label{1}
 z_{n+1} = \lambda - z_{n}^{2},
\end{equation}
where the dynamical variable $z$ and the parameter $\lambda$ are
both complex. The set of parameter values defined by the condition
that the iterations launched from the critical point $z=0$ do not
diverge to infinity is the celebrated Mandelbrot set, perhaps, the
most well-known example of a fractal~\cite{Peitgen,Devaney}. Among
other interesting objects in the field of complex analytic
dynamics one can mention Julia sets -- fractal basin boundaries of
the attractor at infinity on the $z$-plane~\cite{Peitgen,Devaney},
period-tripling and other unusual bifurcation
phenomena~\cite{Golberg,Cvitanovic1,Cvitanovic2}, Siegel discs --
domains on $z$-plane, filled by closed invariant curves, which
appear near fixed points at the moment of stability loss via
irrational eigenvalues~\cite{Widom,Manton,MacKay}. Although some
nontrivial physical applications of complex maps are known (for
problems like renormalization group approach in phase-transition
theory and percolation theory~\cite{Hu,percol}), it would be
interesting to find examples of nonlinear systems manifesting one
or more of the above mentioned phenomena in actual dynamical
behavior.

In fact, complex analytic functions represent a very special and
restricted class of maps. Indeed, the real and imaginary parts of
$f(z)$ must satisfy the Cauchy - Riemann equations. If this is not
the case, the dynamics become drastically
different~\cite{Peinke,Klein,Peckham}. This circumstance forces
one to ask the principal question: Do the phenomena demonstrated
by complex analytic maps have any concern to dynamical behavior of
physical systems? Recently this problem was posed and discussed by
Beck~\cite{Beck}. This author has considered motion of a particle
in a double-well potential in a time-dependent magnetic field, and
proved that under certain assumptions a complex analytic map may
describe the dynamics of the particle.

The aim of the present paper is to suggest some general approach
to constructing physical systems with properties of dynamics
specific to the complex analytic maps.

Let us start from the complex quadratic map ~(\ref{1}) and, first,
separate the real and imaginary parts. Designating $z=x+iy$ and
$\lambda=\lambda'+i\lambda''$ we obtain
\begin{equation}\label{2}
x_{n+1} = \lambda' - x_{n}^{2} + y_{n}^{2}, \qquad y_{n+1} =
\lambda''- 2 x_{n} y_{n}.
\end{equation}
The variable and parameter change
\begin{equation}\label{3}
\xi = x + \beta y, \qquad \eta = x - \beta y,
\end{equation}
\begin{equation}\label{4}
\lambda_{1} = \lambda' + \beta \lambda'', \qquad \lambda_{2} =
\lambda' - \beta \lambda'',
\end{equation}
where $\beta \neq 0$ is an arbitrary constant, transforms the
equations~(\ref{2}) into a set of two coupled real logistic maps
\begin{equation}\label{5}
  \xi_{n+1} = \lambda_{1} - \xi_{n}^{2} + \varepsilon (\xi_{n} -
\eta_{n})^{2}, \quad
  \eta_{n+1} = \lambda_{2} - \eta_{n}^{2} +
\varepsilon (\xi_{n} - \eta_{n})^{2}.
\end{equation}
Here $\varepsilon = (1 + \beta^{2})/4 \beta^{2}$ plays the role of
a coupling parameter. It is worth noting that coupling in these
equations is of a very special kind: It may be interpreted as an
equal simultaneous shift of control parameters in both maps at
each step of iterations, which is proportional to the square of
the dynamical variable difference.

It is easy to see that for any selection of $\beta$ the coupling
parameter is positive, and exceeds $1/4$. Nevertheless, we may
consider the coupled maps~(\ref{5}) for arbitrary values of
$\varepsilon$. To motivate this, we turn to the generalized
complex numbers~\cite{Lavrentjev,perplex1,perplex2}. For pairs of
real numbers $(x,y)$ written as $x+iy$ it is possible to define
different consistent rules of arithmetical operations setting
$i^{2} = a+ib$, where $a$ and $b$ are some real constants. The
case $a=-1$, $b=0$, or $i^{2}=-1$, gives rise to usual complex
numbers, $a=1$, $b=0$, $i^{2}=1$ -- to the so-called perplex
numbers, and $a=b=0$, $i^{2}=0$ -- to dual numbers. Any other
selection of $a$ and $b$ appears to be isomorphic to one of these
three cases, which are known as elliptic, hyperbolic and parabolic
number systems, respectively.

In the case of perplex numbers instead of Eqs.~(\ref{2}) we have
$x_{n+1} = \lambda' - x_{n}^{2} - y_{n}^{2}$, $y_{n+1} = \lambda''
-2 x_{n} y_{n}$. Then the variable change~(\ref{3}),~(\ref{4})
yields Eq.~(\ref{5}) with $\varepsilon = (\beta^{2}-1)/4
\beta^{2}$. The coupling parameter can take either positive or
negative values satisfying $\varepsilon<1/4$. (In fact, by an
appropriate variable change this case can be reduced to two
uncoupled real maps.) For dual numbers $x_{n+1} = \lambda' -
x_{n}^{2}$, $y_{n+1} = \lambda'' - 2 x_{n} y_{n}$. By means
of~(\ref{3}),~(\ref{4}) we arrive at the coupled maps~(\ref{5})
with $\varepsilon = 1/4$ -- the same value of the coupling
parameter for any $\beta$.

Diagrams in Fig.~1 are charts of the parameter plane
$(\lambda_{1},\lambda_{2})$ for the coupled maps~(\ref{5}) for
several values of the coupling parameter. To obtain them we
produce a large number of iterations starting from the origin $\xi
= \eta = 0$ at each pixel of some area on the parameter-plane and
analyze the asymptotic behavior of the iterations. Divergence is
marked by white, aperiodic behavior by black, and asymptotically
periodic dynamics by gray; respective numbers designate the
periods. At $\varepsilon = 0.5$ we observe exactly the Mandelbrot
set (rotated by $45^{\circ}$ in comparison with its usual
depiction). For $0.25 < \varepsilon < +\infty$ this set continues
to exist, but as a distorted version of the standard picture. At
$\varepsilon = 0.25$ the set corresponding to the confined
dynamics turns into a number of strips. For $\varepsilon < 0.25$
it takes the form of rhombus-like structures (at $\varepsilon = 0$
this is a square). Similar metamorphoses from the Mandelbrot set
to the rhombus-like object was noted earlier in
Ref.~\cite{perplex2} for the quadratic map~(\ref{1}) considered
for the generalized complex numbers.

It is known that a single real logistic map represents a
universality class, which is associated with the period-doubling
bifurcation cascade and includes many dissipative nonlinear
systems (forced nonlinear oscillators, R\"ossler and Lorenz
equations, etc.)~\cite{Feigenbaum}. It may be thought that taking
two copies of a system, relating to this universality class, and
maintaining the appropriate type of coupling one can arrange the
dynamical behavior characteristic to the complex analytic maps.
(It is supposed that the control parameters for the both
subsystems allow an independent regulation.)

To illustrate the feasibility of this approach we have elaborated
a real electronic system, which reproduces the dynamics of the
coupled logistic maps (\ref{5}).

We start with a specialized analog device suggested in
Ref.~\cite{Chua} to study the dynamics of nonlinear systems
represented by iterative mappings, in particular, by the logistic
map. Our system (see Fig.~2) contains two pairs of the sample-hold
cells (marked by dotted frames and figures $11$, $12$, $21$, $22$
on the diagram); one pair represents a single real quadratic map.
Each of these cells consists of an analog switch and a capacitor.
In the regime of picking a sample the capacitor is linked to the
signal source via the switch, and accepts a charge up to a
definite voltage. At some moment the switch breaks, and then the
voltage on the capacitor remains constant -- this is the regime of
holding or storage. The voltage from the capacitor governs the
operational amplifier of large input and low output resistance, so
the charge of the capacitor remains practically constant. The used
regime of the operational amplifier ensures equality of the output
voltage to the input one. A set of two sample-hold cells is
governed by two sequences of non-overlapping rectangular pulses:
the switches $K_{11}$ and $K_{21}$ are opened while $K_{12}$ and
$K_{22}$ are closed and vise versa. Multipliers $N_{1}$ and
$N_{2}$ ensure quadratic nonlinearity to obtain squared values of
the voltages corresponding to $\xi_{n}^{2}$, $\eta_{n}^{2}$.
Output of the operational amplifier $D$ is the difference signal
voltage. It is squared by the multiplier $N_{12}$ to produce the
signal $(\xi_{n} - \eta_{n})^{2}$, then multiplied by
$\varepsilon$ and added to the control voltages corresponding to
the parameters $\lambda_{1}$ and $\lambda_{2}$ by means of the
operational amplifiers $L_{1}$ and $L_{2}$. The presence of three
variable resistors $R_{1}^{\sim}$, $R_{2}^{\sim}$, $R_{12}^{\sim}$
gives a possibility to regulate parameters $\lambda_{1}$,
$\lambda_{2}$, and $\varepsilon$, respectively.

Using an oscilloscope in the experiment we could distinguish
either dynamics taking place in a restricted domain of the
voltages, or the voltages jump to some distant values (analog of
divergence in the mathematical model). For periodic regimes the
periods could be easily determined (in units of the period of
pulses, which control the switches) from the picture on the
oscilloscope screen.

Fig.~3 presents two examples of the topography of the plane of two
parameters, namely, of voltages regulated by the variable
resistors $R_{1}^{\sim}$ and $R_{2}^{\sim}$. The values of the
coupling constant are chosen to be $\varepsilon = 0.1$ and
$\varepsilon = 0.5$ respectively.

The first diagram demonstrates a rhombus-like structure, similar
to that in fig.~1d. Some peculiarities of the experimental plot
distinct from the computer model may be explained by inevitable
deflections from perfect quadratic nonlinearities at the edges of
the working interval of voltages.

The second diagram for larger coupling shows a formation
remarkably similar to the Mandelbrot set. Although in the
experiment it was not possible to resolve extremely fine details
of the structure, the location of all main leaves of the "cactus"
are in excellent agreement with the computer generated picture.

Experimental measurements at different values of coupling confirm
that the set on the parameter plane $(U_{1},U_{2})$ corresponding
to the confined dynamics evolve in accordance with the results of
numerical computations for the coupled maps~(\ref{5}).

It is worth emphasizing that what we deal with is a real physical
object, effected by such factors as noise and technical
fluctuations of voltages. The elements are not perfectly
identical, the nonlinear function is not perfectly $x^2$, and so
forth. A substantial circumstance is that all these factors do not
destroy the phenomena of complex analytic dynamics, which we
observe.

Perhaps, an electronic system could be built also to realize in a
straightforward way the dynamics of real and imaginary parts of
the complex variable governed by the map (\ref{1}). However, our
approach seems potentially more interesting because we state a
direction for further search for dynamical systems manifesting
behavior similar to that of the complex iterative maps.

As explained, the quadratic map used as a basic element in our
construction, must be regarded as a representative of the wide
universality class, which includes many realistic physical systems
and their mathematical models (associated with the period-doubling
bifurcation cascade)~\cite{Feigenbaum}. Hence, in the case of a
properly arranged coupling between two period-doubling elements of
any nature, one may expect the whole system to demonstrate the
phenomena of complex analytic dynamics.

\vspace{1cm} The authors acknowledge support from RFBR (grant No
00-02-17509) and from CRDF (REC-006). V.I.P. acknowledges support
from RFBR (grant No 99-02-17735). O.B.I. acknowledges support from
RFBR (grant No 01-02-06385). We thank Carsten Knudsen for
discussion and useful comments.

\newpage
\begin{figure}
\centerline{\includegraphics[scale=1.5]{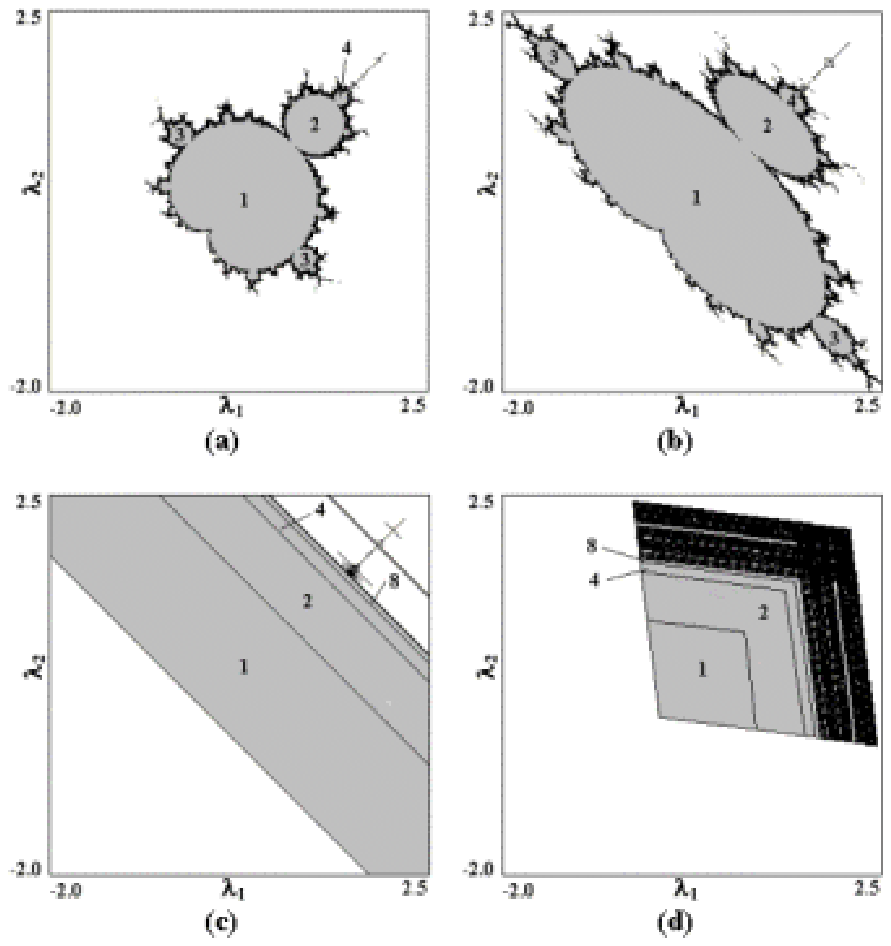}}
\caption{Charts of the parameter plane  for the coupled
maps~(\protect\ref{5}) in dependence on the coupling parameter:
(a) $\varepsilon = 0.5$, (b) $\varepsilon = 0.3$, (c) $\varepsilon
= 0.25$, (d) $\varepsilon = 0.1$. Divergence is marked by white,
aperiodic behavior by black, and asymptotically periodic dynamics
by gray; periods are shown by respective numbers.}
\end{figure}

\begin{figure}
\centerline{\includegraphics[scale=1.2]{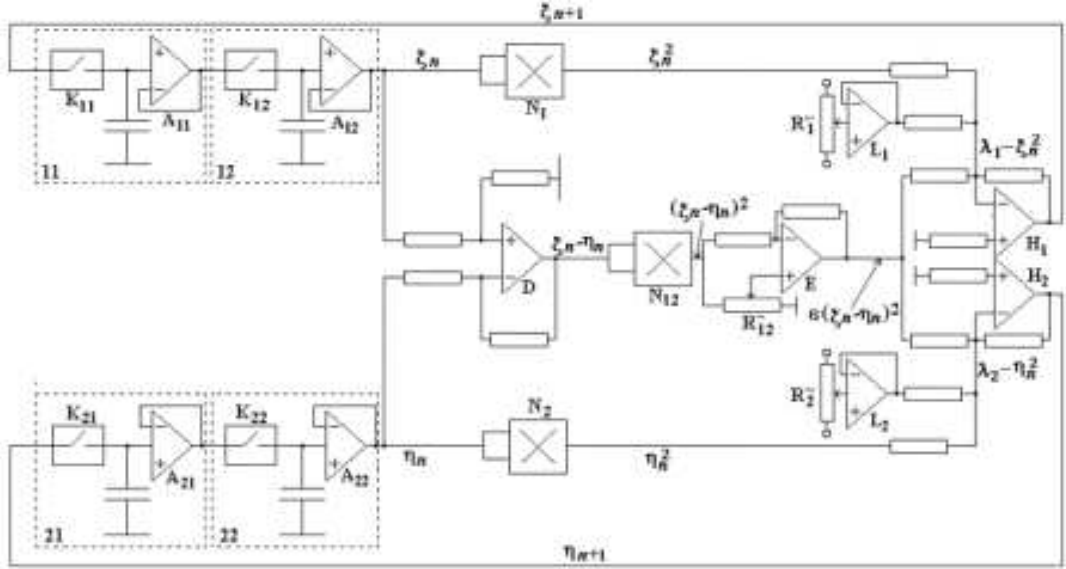}}
\caption{Schematic representation of the electronic device
corresponding to the coupled maps~(\protect\ref{5}). Dashed frames
show the sample-hold cells. $K_{11}$, $K_{12}$, $K_{21}$, $K_{22}$
are the electronic switches controlled by sequences of the
rectangular pulses. $A_{11}$, $A_{12}$, $A_{21}$, $A_{22}$, $D$,
$E$, $L_{1}$, $L_{2}$, $H_{1}$ and $H_{2}$ are the operational
amplifiers, $N_{1}$, $N_{2}$, and $N_{12}$ are the multipliers.}
\end{figure}

\begin{figure}
\centerline{\includegraphics[scale=1.5]{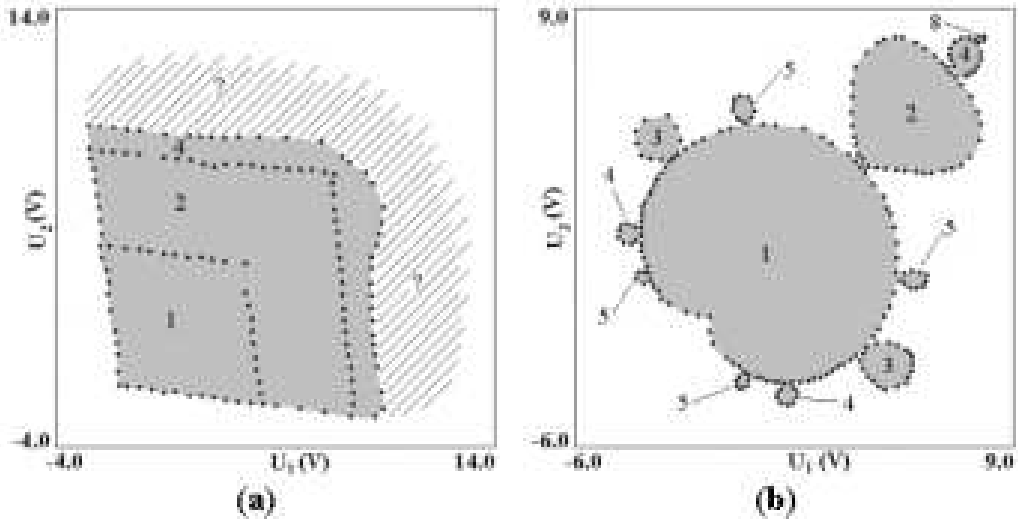}}
\caption{Configuration of the set corresponding to dynamics in a
restricted domain in the experiment with the electronic circuit.
The chart represent the plane of voltages $(U_{1},U_{2})$ with
$U_{1} \simeq 5 \lambda_{1}$, $U_{2} \simeq 5 \lambda_{2}$
controlled by variable resistors $R_{1}^{\sim}$ and
$R_{2}^{\sim}$. The coupling parameter values are $\varepsilon =
0.1$ (a) and $\varepsilon = 0.5$ (b). The hatching on diagram (a)
marks the domains of complex behavior}
\end{figure}

\end{document}